\documentclass[10pt]{iopart}
\usepackage{graphicx}
\usepackage{color}
\usepackage{amssymb}
\usepackage{iopams}
\usepackage[%
  breaklinks=true,
  colorlinks=true,
  linkcolor=blue,
  urlcolor=blue,
  citecolor=blue,
]{hyperref}
\begin{document}
\title{Long range of indirect exchange interaction on the edges of MoS$_{2}$ flakes}

\author{O \'Avalos-Ovando$^{1,*}$, D Mastrogiuseppe$^{2}$, and S E Ulloa$^{1}$}

\address{$^1$ Department of Physics and Astronomy, and Nanoscale and Quantum
	Phenomena Institute, \\ Ohio University, Athens, Ohio 45701--2979, USA}
\address{$^2$ Instituto de F\'isica Rosario (CONICET), 2000 Rosario, Argentina}

\ead{oa237913@ohio.edu}

\begin{abstract}
We study the Ruderman-Kittel-Kasuya-Yosida interaction between two magnetic impurities connected to the edges of zigzag-terminated MoS$_{2}$ flakes. When the impurities lie on the edges of the flake, the effective exchange interaction exhibits sizable noncollinear Dzyaloshinskii-Moriya character that competes with a strong Ising coupling. We analyze the characteristic decay exponent for doping levels inside the band gap of the infinite layer, corresponding to edge states of the flake at the Fermi level. The characteristic exponents show sub-two-dimensional (sub-2D) behavior for these band fillings, with decays much slower than quadratic. The Ising interaction has effectively one-dimensional (1D) long range, while the noncollinear component that grows for short impurity separation becomes comparable in magnitude. The resulting tunable exchange interaction on these systems opens the way for the study of interesting phases of impurity arrays with long-range stable helical order.
\end{abstract}


\maketitle
\ioptwocol

\section{Introduction}
Layered transition-metal dichalcogenides (TMDs) \cite{Novoselov2005,Geim2013} have been receiving great attention due to their unique electronic and optical properties. These materials can be exfoliated down to a stack of three atomic layers, in which one layer of transition metal atoms is sandwiched between two layers of chalcogen species. Each material (e.g. MoS$_2$, WS$_2$, among others) has different electronic structure and properties, such as strong photoluminescence, large spin-orbit coupling (SOC) \cite{Zhu2011,Cheiwchanchamnangij2012,Xiao2012}, and either direct or indirect band gap depending on the number of layers \cite{Mak2010}. These properties make TMDs very promising for applications and devices \cite{Radisavljevic2011}.

One important topic of study that remains somewhat unexplored, however, is the presence of magnetic dopants in TMDs and how they interact among each other, giving rise to interesting magnetic properties driven by the TMD substrate. The combination of strong SOC and the presence of magnetic dopants could render these materials good candidates for spintronics applications \cite{Han2016}. The magnetic dopants, or impurities, can be incorporated during the sample production process or can be introduced extrinsically afterwards.
Recently, substitutional Mn \cite{Zhang2015,Wang2016,Huang2017,Tan2017}, Cr \cite{Huang2017} and Co \cite{Liu2017NatChem,Nethravathi2017} impurities have been successfully incorporated in experiments with MoS$_2$.
Density functional theory (DFT) studies have proposed that impurities of Mn \cite{Mishra2013,Andriotis2014,Lu2014,Cong2015}, Fe \cite{Mishra2013,Andriotis2014,Lu2014,Cong2015,Shu2015}, Co \cite{Mishra2013,Andriotis2014,Lu2014,Wang2016electronic}, Ni \cite{Mishra2013}, Cr \cite{Vahakangas2017}, V \cite{Vahakangas2017} and several other transition metals \cite{Cheng2013}, would be stable when hybridized into TMD monolayers.
On the other hand, impurities such as Cr, Mn, Fe and Co, have been proposed to be stable near the {\em edges} of a monolayer, with either metal or sulfur termination  \cite{Saad2016}.
And very recent experiments have shown that Co impurities bind to vacancies at MoS$_{2}$ sulfur-edges \cite{Liu2017NatChem,Nethravathi2017}. Most of these recent predictions and experimental observations of magnetic dopants show a rapidly growing trend of research with a promising future for possible spintronics applications. Therefore, it becomes important to understand the interaction between the electronic degrees of freedom and localized magnetic moments embedded in TMDs.
We focus here on the role that robust states localized near flake edges in these materials play, providing effective long range interactions between such magnetic impurities hybridized near crystallite borders.

Magnetic species in a metallic sample can interact effectively via the conduction electrons by the Ruderman-Kittel-Kasuya-Yosida (RKKY) interaction \cite{RudermanKittel1954,Kasuya1956,Yosida1957}.  In conventional metals, the interaction can be written as $\propto [2 k_F r]^{d} \cos{(2 k_F r)}$ \cite{Fischer1975,Litvinov1998}, the superposition of an oscillatory component and a decaying envelope ($[2 k_F r]^{d}$, $d<0$), where $r$ is the distance between impurities, $k_{F}$ is the Fermi momentum and $|d|$ is the dimensionality of the host electron system. The first component describes how the interaction changes between ferromagnetic (FM) and antiferromagnetic (AFM) alignment of the magnetic moments
(with characteristic scale given by half the Fermi wavelength).
However, the spin and orbital content of the conduction states involved can result in unusual interaction features.
For instance, in graphene, the interaction is found to decay as $r^{-3}$ at the Dirac point, while in doped or spin-polarized graphene it decays as $r^{-2}$ \cite{Parhizgar2013graphene}, as in conventional 2D materials. As we will see, TMD edges result in unusual decay with
$|d|<2$, which we call \emph{sub-2D} behavior.

Many TMDs are semiconducting materials, but can be doped with electrons and/or holes. This allows one to reach energy sectors in which the SOC has strong influence in the band structure. It is well known that SOC can introduce anisotropic interactions between magnetic moments \cite{Imamura2004}, which can give rise to helical spin assembly structures.
In TMDs the band splitting due to SOC is much larger near the valence band maximum (VBM) than close to the conduction band minimum (CBM), so that the \emph{p}-doped case is more attractive for studying the RKKY interaction between localized magnetic impurities, and find possible ways of tuning the interaction for further use.
The RKKY interaction on  \emph{p}-doped TMD 2D-bulk monolayers has been studied in previous works, considering only the bands near the Brillouin zone corners $K$ and $K'$ \cite{Parhizgar2013,Hatami2014}, and with the inclusion of the important valence band at the $\Gamma$ point, whose maximum lies close in energy to those at the $K,K'$ valleys \cite{Mastrogiuseppe2014}.

Although analyzing the case of infinite layers is essential, the process of exfoliation often produces nanoscale samples with different shapes--often triangular--and boundaries \cite{Chiu2014,Van2013,Lauritsen2007}. Finite samples often have richer properties than the bulk counterparts, with characteristics that depend on the shape and termination.
For instance, MoS$_{2}$ zigzag nanoribbons exhibit unusual ferromagnetic properties \cite{Li2008,Botello2009,Tongay2012}, probably due to the presence of edge states \cite{Bollinger2001}. Intrinsic magnetism also arises in triangular MoS$_{2}$ films, depending on the type of edge termination \cite{Lauritsen2007,Zhang2007}. TMD zigzag edges have also been proposed to host Majorana fermions when in proximity to superconductors  \cite{Xu2014,Chu2014}, and MoS$_{2}$ triangular flake edges to host collective plasmon-like excitations with strong optical response \cite{Rossi2017}.

Using an effective three-orbital tight-binding model \cite{Liu2013} for MoS$_{2}$, we present here a systematic study of the interaction decay between  magnetic impurities along the edges of flakes, where carriers in such localized states mediate the interaction.  We analyze the features of the exchange, including the appearance of a noncollinear component due to the strong SOC, and focus on the decay exponent of the interactions as the chemical potential sweeps the midgap states, reached experimentally by gating or doping \cite{Van2013,Lu2014v2}. We find that the exponent signals general sub-2D behavior, with the Ising component dominating over the in-plane components, especially at larger separations.  Moreover, the noncollinear Dzyaloshinskii-Moriya (DM) component increases gradually with separation, before reaching a strength the competes with the Ising interaction.  Although we had found 2D behavior for the interaction between impurities in the interior of flakes before \cite{Avalos2016v2}, and some evidence of long range interaction for hybridization near the edges \cite{Avalos2016}, we present here the full systematics.  We find that indeed whenever the Fermi level is midgap, the sub-2D behavior dominates the effective exchange interaction, with unusual long-range characteristics. These results suggest that whenever magnetic impurities lie at/near the edges of TMD flakes, the long-ranged interaction with strong noncollinear character may give rise to interesting helically ordered impurity assemblies with stable phases.  As the Fermi level and impurity separation may be tunable via doping and manipulation, respectively, these systems may provide interesting new playground to explore the behavior of 1D magnetic ensembles. Although such helical structures have been recently shown controllable in metallic systems \cite{Zhou2010,Khajetoorians2016,Steinbrecher2017}, our predictions on TMDs would have interesting advantages, since several TMDs have similar band gaps, large SOC strengths, and well-defined crystallite edges, suggesting the exploration of long-range interaction would be supported by stable and accessible structures, less susceptible to quantum or thermal fluctuations.

\section{Model and Approach}

\begin{figure}[ht]
   \centering
   \includegraphics[width=0.45\textwidth]{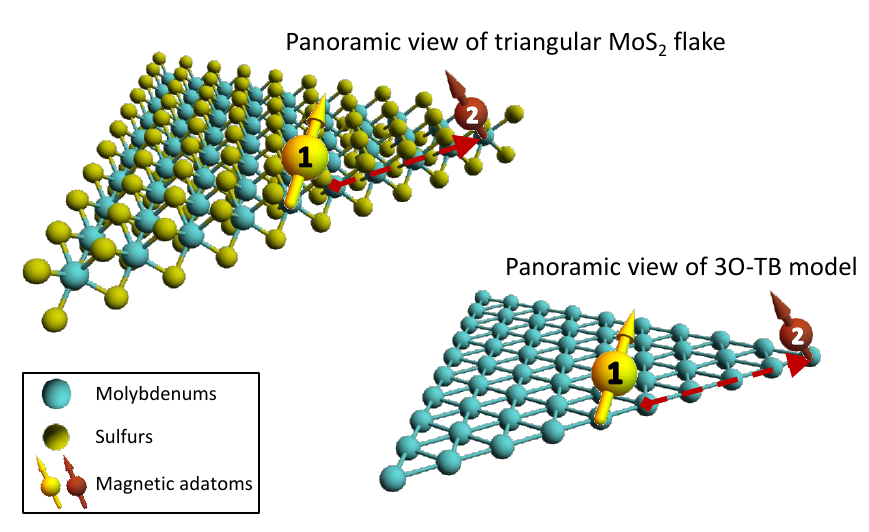}\\
   \caption{\label{fig1} Views of the flake and effective tight-binding model with  three Mo $d$-orbitals  considered. Two magnetic impurities  are connected to the edge of the flake, shown as yellow and brown arrows. Impurity $1$ is held fixed while impurity $2$ changes location along the zigzag edge,  as shown by the dashed-brown trajectory.}
\end{figure}

We consider triangular zigzag-terminated MoS$_{2}$ nanoflakes \cite{Van2013}, with two magnetic impurities hybridized to different lattice sites, as shown
in figure\ \ref{fig1}.
A suitable approximation is to describe the system by a lattice of only Mo atoms (see figure \ref{fig1}) since at low energies (near the optical gap) three Mo \emph{d}-orbitals ($d_{z^2}$, $d_{xy}$ and $d_{x^2-y^2}$) contribute the most to the states, with almost no $p$-orbital weight from the chalcogens. The Hamiltonian is given by $H = H_{\mathrm{3OTB}} + H_{\mathrm{I}}$,
where $H_{\mathrm{3OTB}}$ describes the triangular TMD lattice and $H_{\mathrm{I}}$ is the Hamiltonian for the magnetic impurities interacting with the charge carriers in the host. The MoS$_{2}$ tight-binding Hamiltonian can be written as $H_{\mathrm{3OTB}} = H_{\mathrm{o}} + H_{\mathrm{t}} + H_{\mathrm{SOC}}$, with
\begin{equation}\label{lattice2}
  H_{\mathrm{o}} = \sum_{ \mathbf{l}}^{N_{sites}} \sum_{s=\uparrow,\downarrow} \sum_{\alpha,\alpha'} \varepsilon_{\alpha\alpha',s}d_{\alpha,\mathbf{l},s}^{\dagger}d_{\alpha',\mathbf{l},s},
\end{equation}
where $d_{\alpha,\mathbf{l},s}$ ($d^{\dagger}_{\alpha,\mathbf{l},s}$) annihilates (creates) a spin-$s$ electron in orbital $\alpha$ $\in\,\left\{d_{z^2},d_{xy},d_{x^2-y^2}\right\}$ on site $\mathbf{l}=l_{1}\mathbf{a}_{1}+l_{2}\mathbf{a}_{2}$; $\mathbf{a}_{1}=a(1,0)$, $\mathbf{a}_{2}=a(1/2,\sqrt{3}/2)$ are lattice vectors of the triangular Mo lattice, with $a=3.19$ \AA. The nearest-neighbor hopping Hamiltonian is given by
\begin{equation}
   H_{\mathrm{t}} = \sum_{\mathbf{l,a}_j} \sum_{s=\uparrow,\downarrow} \sum_{\alpha,\alpha'}
   t_{\alpha\alpha'}^{(\mathbf{a}_{j})}d_{\alpha,\mathbf{l},s}^{\dagger}d_{\alpha',\mathbf{l}+\mathbf{a}_{j},s}+
   \mathrm{H.c.},\\
\end{equation}
where $t_{\alpha\alpha'}^{(\textbf{a}_{j})}$ are hopping parameters in three different directions, $j=1,2,3$, with $\textbf{a}_3=\textbf{a}_2 - \textbf{a}_1$. There are 27 hopping parameters, 9 for each direction. For direction $\textbf{a}_{1}$, they are taken from \cite{Liu2013} (Table II), while for directions
$\textbf{a}_{2}$ and $\textbf{a}_{3}$ they are calculated from symmetry transformations (see e.g. Eqs.\ (4-6) in \cite{Pavlovic2015}). The SOC is approximated by considering the onsite Mo contributions as $H_{\mathrm{SOC}}=\lambda L_{z}S_{z}$, where $L_{z}$ is the $z$ component matrix of the orbital angular momentum, $S_{z}$ is the spin Pauli matrix, and $\lambda$ is the SOC strength. This results in different orbital mixing within the same site,  $\varepsilon_{d_{xy}d_{x^2-y^2},\uparrow}=\varepsilon_{d_{x^2-y^2}d_{xy},\downarrow}=i\lambda$ and $\varepsilon_{d_{xy}d_{x^2-y^2},\downarrow}=\varepsilon_{d_{x^2-y^2}d_{xy},\uparrow}=-i\lambda$. We use $2\lambda=150$ meV, in good agreement with DFT calculations \cite{Xiao2012,Liu2013} and experimental values (152 meV \cite{Sun2013}, 138 meV \cite{Alidoust2014} and $145$ meV \cite{Miwa2015}).
With this Hamiltonian and the right choice of boundaries, one can construct a flake like the triangle we consider \cite{Avalos2016,Avalos2016v2,Pavlovic2015}. Others TMDs can be similarly modeled, taking the onsite, hopping and SOC parameters from Tables II and III in \cite{Liu2013}.

\begin{figure}[ht]
   \centering
   \includegraphics[width=0.45\textwidth]{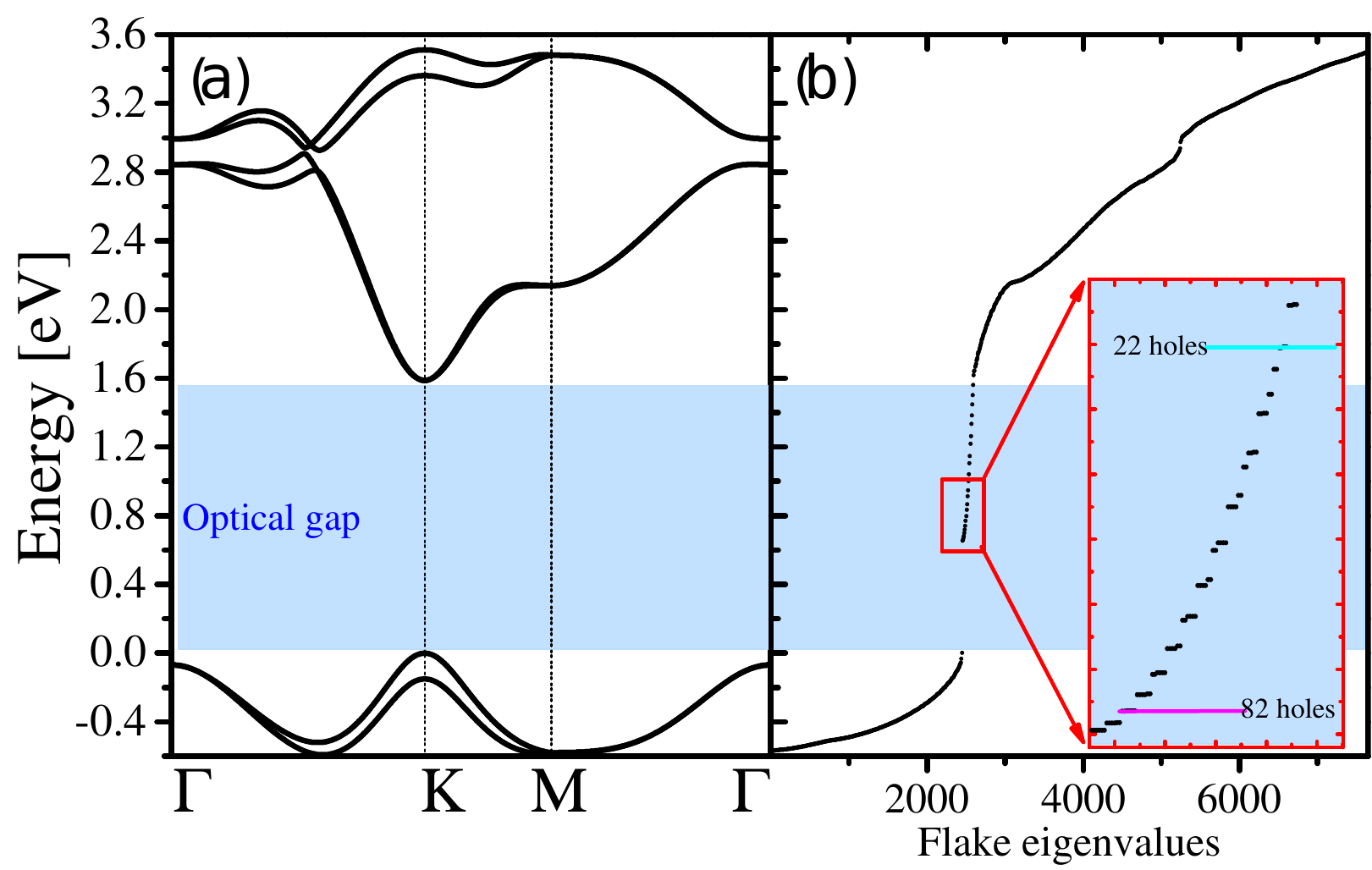}\\
   \caption{\label{fig2}(a) Band structure for bulk MoS$_{2}$ monolayer in the 3OTB model. The gap is shown as a light blue area. The VBM at $K$ is shifted to zero energy. (b) Discrete energy levels for a 50-row flake, showing midgap energy levels. Midgap state typical distribution is shown in the red inset, where degeneracies due to spin and threefold symmetry are evident.}
\end{figure}

In the infinite 2D MoS$_{2}$ monolayer, the first Brillouin zone  has two inequivalent $K$ and $K'$ points, with a sizable spin-splitting around the VBM due to SOC [see figure\ \ref{fig2}(a)]. There is a direct band gap ($\sim1.6$ eV) between the VBM and the CBM at these two points, with definite spin-valley relation due to the absence of inversion symmetry in the unit cell. On the other hand, for finite systems the electronic spectrum [see figure\ \ref{fig2}(b)] shows both bulk- and edge-like states. The latter, coming from both the valence and conduction bands, appear into the 2D bulk gap, signaling the presence of 1D-like extended states localized near the borders of the sample \cite{Bollinger2001,Segarra2016,italians}.

The Hamiltonian for the magnetic impurities connected to specific sites of the TMD lattice is given by
$H_{\mathrm{I}}={\cal J}_\alpha \sum_{i=1,2}\textbf{S}_{i}^{\mathrm{I}}\cdot\textbf{S}_{\alpha}(\textbf{l}_{i})$,
where ${\cal J}_\alpha$ is the exchange coupling between the localized magnetic moment (either $i=1,2$) and the conduction electrons at lattice site $\textbf{l}_{i}$ and orbital $\alpha$ ($d_{z^2},d_{xy}$ or $d_{x^2-y^2}$). $\textbf{S}_{i}^{\mathrm{I}}$ is the spin operator for the local magnetic moment $i$, and
  $\textbf{S}_{\alpha}(\textbf{l}_{i})$
is the electron spin operator at site $\textbf{l}_{i}$ for orbital $\alpha$. When ${\cal J}$ is small, the electronic degrees of freedom can be integrated out (using second order perturbation theory, for example) in the bulk 2D crystal,  yielding an effective exchange interaction of the form
\begin{eqnarray}\label{jeffective1}
H_{RKKY} = &J_{XX}\left(S_{1x}^{\mathrm{I}}S_{2x}^{\mathrm{I}}+S_{1y}^{\mathrm{I}}S_{2y}^{\mathrm{I}}\right)+\nonumber\\
&J_{ZZ} \,S_{1z}^{\mathrm{I}}S_{2z}^{\mathrm{I}}+J_{DM}\left(\textbf{S}_{1}^{\mathrm{I}}\times \textbf{S}_{2}^{\mathrm{I}}\right)_{z},
\end{eqnarray}
where the $J$'s are proportional to the static spin susceptibility tensor of the electron gas \cite{RudermanKittel1954,Kasuya1956,Yosida1957}. The total effective interaction is a competition between in-plane $J_{XX}(=J_{YY})$, Ising $J_{ZZ}$, and Dzyaloshinskii-Moriya  $J_{DM}$ terms. In TMDs the Ising and DM terms are generated by the SOC; their competition has been recently discussed for bulk samples \cite{Parhizgar2013,Mastrogiuseppe2014} and flakes \cite{Avalos2016,Avalos2016v2}.
To calculate the effective $J$'s in our finite sample, we take the difference of triplet and singlet configurations of impurities in the electronic ground state \cite{Deaven1991,Black2010} as
\begin{equation}\label{jeffective2}
   J_{\beta\beta'}^{\alpha\alpha'}=\frac{E(\uparrow_{\beta},\uparrow_{\beta'})-E(\uparrow_{\beta},\downarrow_{\beta'})}{2S^{2}}.
\end{equation}
with $S=1/2$ in our case. $\beta$, $\beta'$ represent the spin projection ($x, y, z$) for magnetic impurities. For instance, $J_{XY}^{d_{z^2}d_{x^2-y^2}}$ ($XY=DM$ throughout the text) is the interaction strength between impurities when the spin of the first points along the $x$ direction and is hybridized to a $d_{z^2}$ orbital in the lattice, while  the spin of the second points along $y$ and is hybridized to a $d_{x^2-y^2}$ orbital. This non-perturbative approach has been proved to be valid even for large values of local ${\cal J}_\alpha$ and is capable of generating results for any separation between impurities \cite{Black2010}. The ground state energy of the system, including both impurities, is calculated from the eigenvalue spectrum of the full Hamiltonian up to the Fermi energy $\epsilon_{\mathrm{F}}$, as $E(\textbf{S}^{\mathrm{I}}_{1},\textbf{S}^{\mathrm{I}}_{2})=\sum_{\epsilon_{i,s}\leq \epsilon_{\mathrm{F}};s=\uparrow,\downarrow} \epsilon_{i,s}  $, where the $\epsilon_{i,s}$ are obtained by exact numerical diagonalization of the full Hamiltonian, given here by a matrix of $6 N_{sites}\times 6 N_{sites}$.

\section{Results}
We present here results for triangular flakes of 50 rows, with a total of 1275 Mo atoms ($\approx160$ \AA\ on edge), within the range of sizes of real  flakes that go from just a few sites \cite{Lauritsen2007} to several micrometers on edge \cite{Van2013}. This  flake, large enough to yield experimentally relevant results, is shown schematically in figure\ \ref{fig1}. We study the RKKY interaction for doping levels lying inside the optical gap [see figure\ \ref{fig2}(b)].  In this case,  nearly 100 energy states have been brought from the bulk bands into the gap, with dominating $d_{z^2}$ character. As such, midgap levels correspond to $p$-dopings from $1.82\times10^{12}$ to $9.08\times10^{13}$ holes/cm$^{2}$. Such MoS$_{2}$ \emph{p}-doping can be achieved by substituting Mo with Nb \cite{Laskar2014,Suh2014}, and are also predicted for different dopants \cite{Mishra2013,Cheng2013,Dolui2013}, including P implantation \cite{Nipane2016}.

Two magnetic impurities are hybridized to an edge of the flake, keeping one fixed and moving the second one along the zigzag edge, as shown schematically in figure\ \ref{fig1}. We fix the first impurity at the 10th row from the flake tip (yellow impurity $1$ in figure\ \ref{fig1}), and move the second one from the 11th row onwards (brown impurity $2$ in figure\ \ref{fig1}). The location of the impurities here is arbitrary but the results exhibit the main qualitative behavior, characterized by the impurity separation. We set ${\cal J}_\alpha=0.3$ eV, and consider both impurities to be hybridized to Mo $d_{z^2}$ orbitals, since they are dominant on the edge states \cite{Avalos2016,Segarra2016}. This value of ${\cal J}_\alpha$ is a choice in accordance with predictions that range from a few meV to a couple hundred meV \cite{Cong2015,Shu2015,Vahakangas2017,Qi2016}, even at the edges\cite{Saad2016}. In general, microscopic calculations could provide input for specific material-impurity pairs.

As mentioned, all midgap levels have wave functions entirely localized at/near the edges of the flake. Notice that energy levels fall into groups of six nearly-degenerate states, with wave function modulations along the edge, due to finite size effects. For instance, the first 6 midgap states closest to the VBM [see inset of figure\ \ref{fig2}(b)] show one antinode located in the middle of the flake edge, while the last 6 midgap states have 15 antinodes with a 3-site period along the flake edge. The interaction $J_{\beta\beta'}^{d_{z^2}d_{z^2}}$ between magnetic impurities hybridized to these modulated wave functions on the edges, is largely dependent on the energy doping \cite{Avalos2016}, since different states close to a given Fermi level carry different information on spin/spatial structure.  This gives rise to complex oscillation patterns on the interaction between impurities.  Effective RKKY interactions between impurities hybridized to other orbitals, $d_{xy}$ and $d_{x^2-y^2}$, are also possible on the edges, but are found to have smaller strength than for $d_{z^2}$ hybridization, given their correspondingly smaller contribution to the states. However, those interactions show the same separation dependence and relative strength of different exchange terms.

\begin{figure*}[ht]
	\centering
	\includegraphics[width=0.5\textwidth]{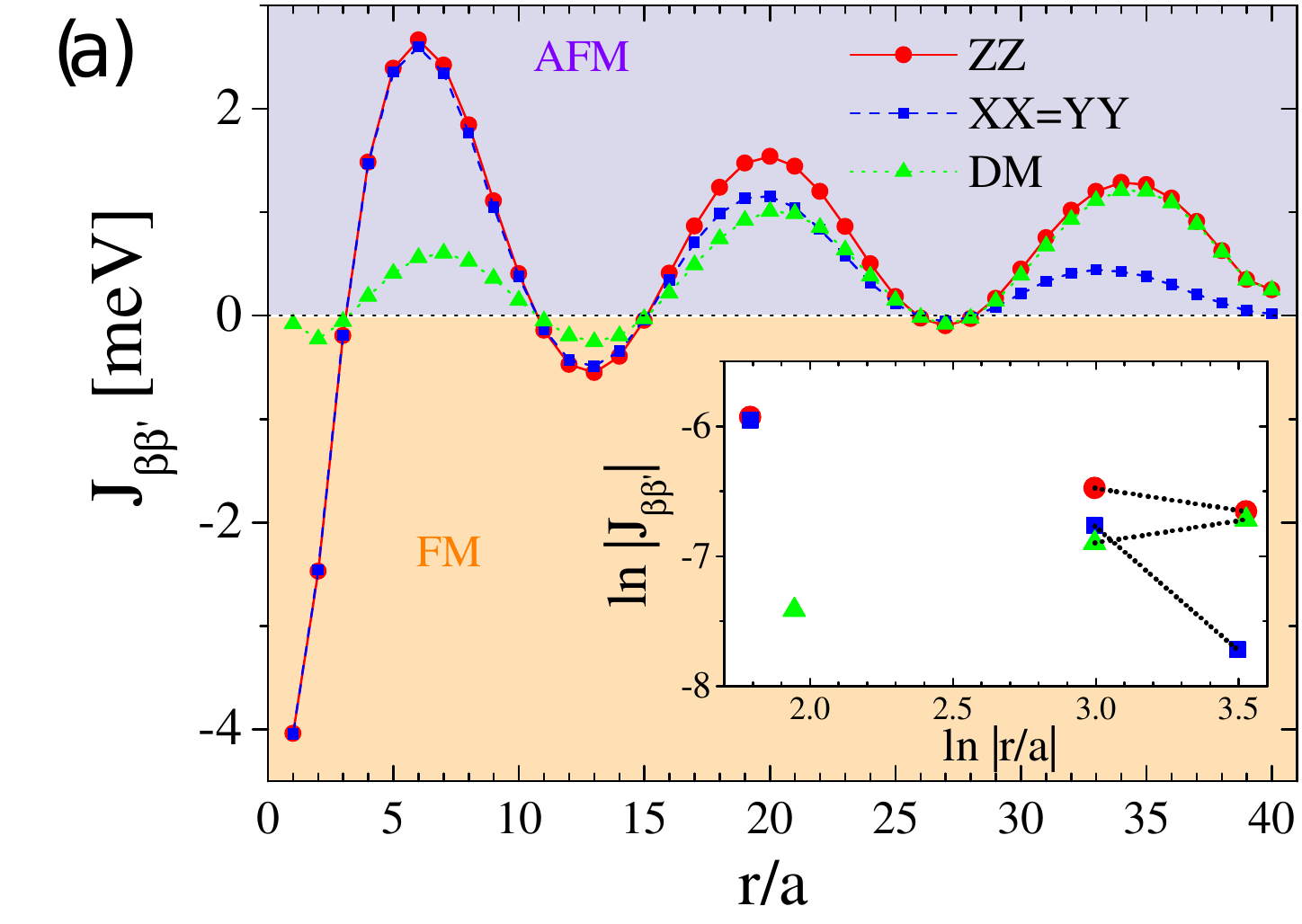}\includegraphics[width=0.5\textwidth]{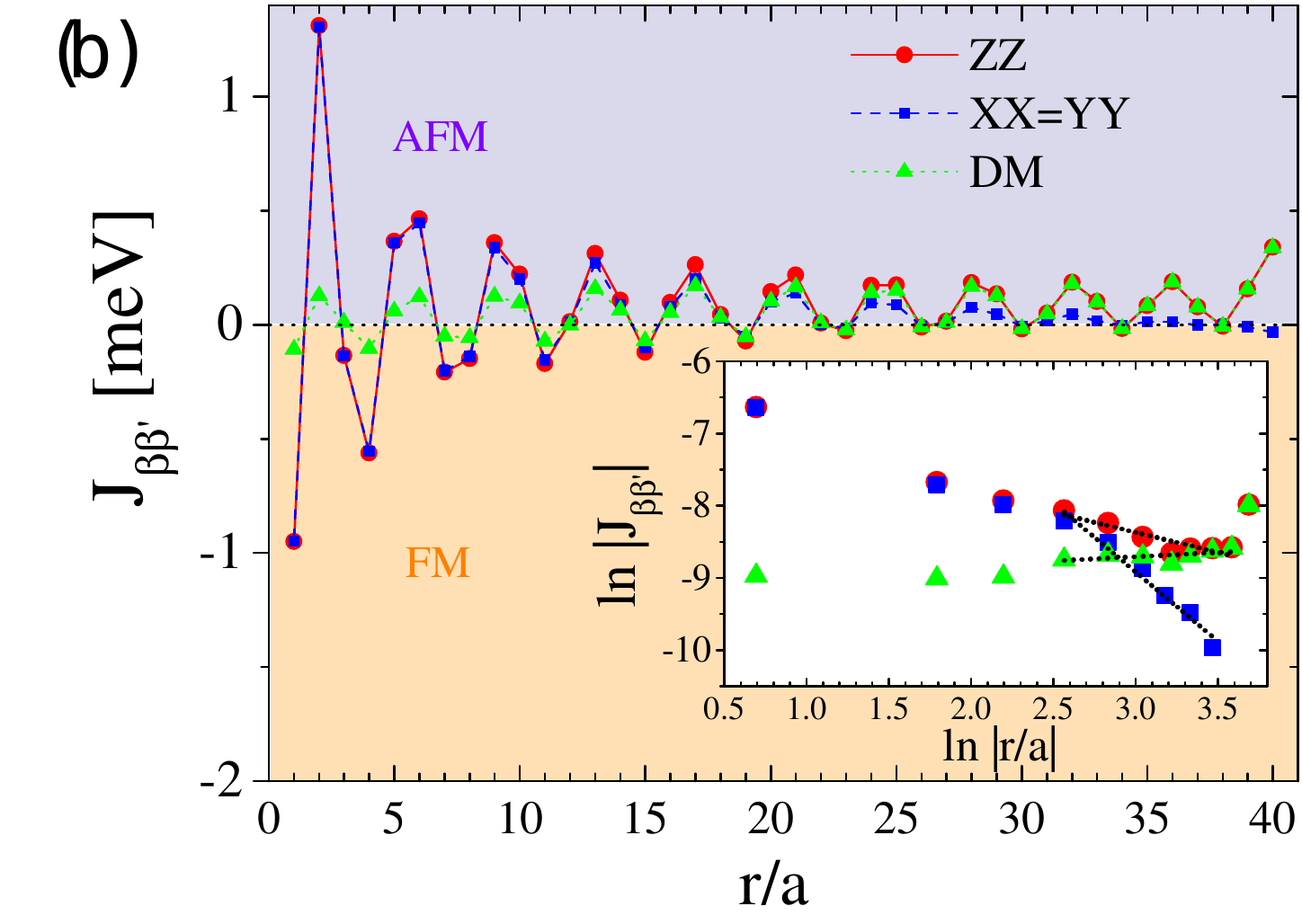}\\
	\includegraphics[width=0.5\textwidth]{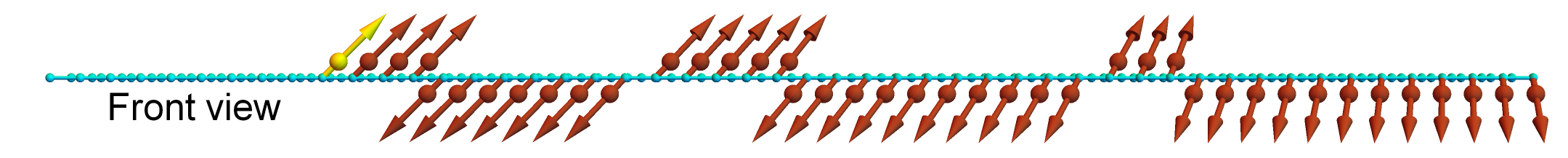}\includegraphics[width=0.5\textwidth]{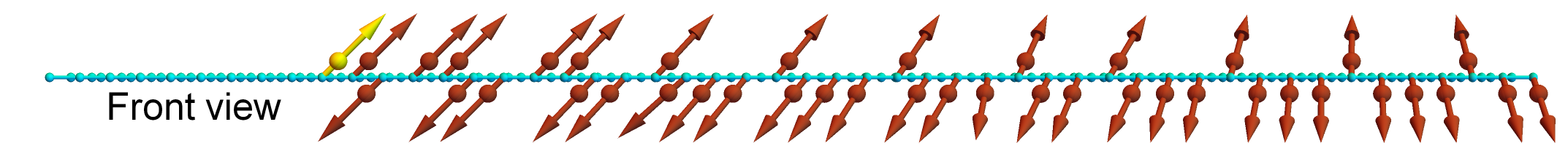}\\
	\includegraphics[width=0.5\textwidth]{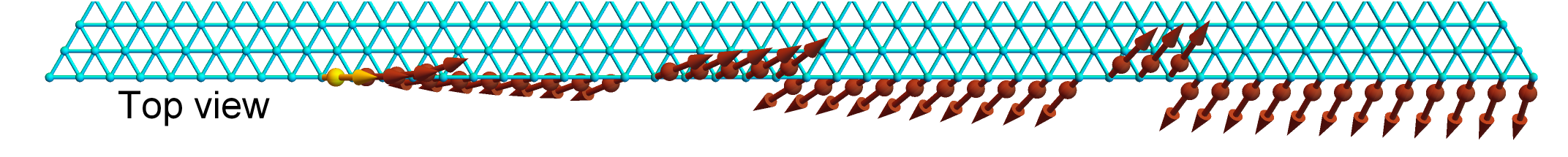}\includegraphics[width=0.5\textwidth]{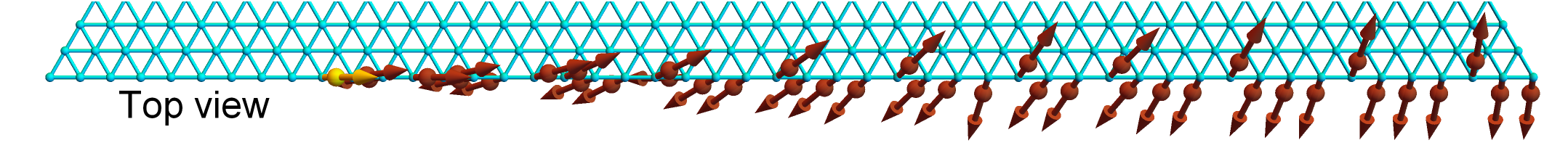}\\
	\includegraphics[width=0.5\textwidth]{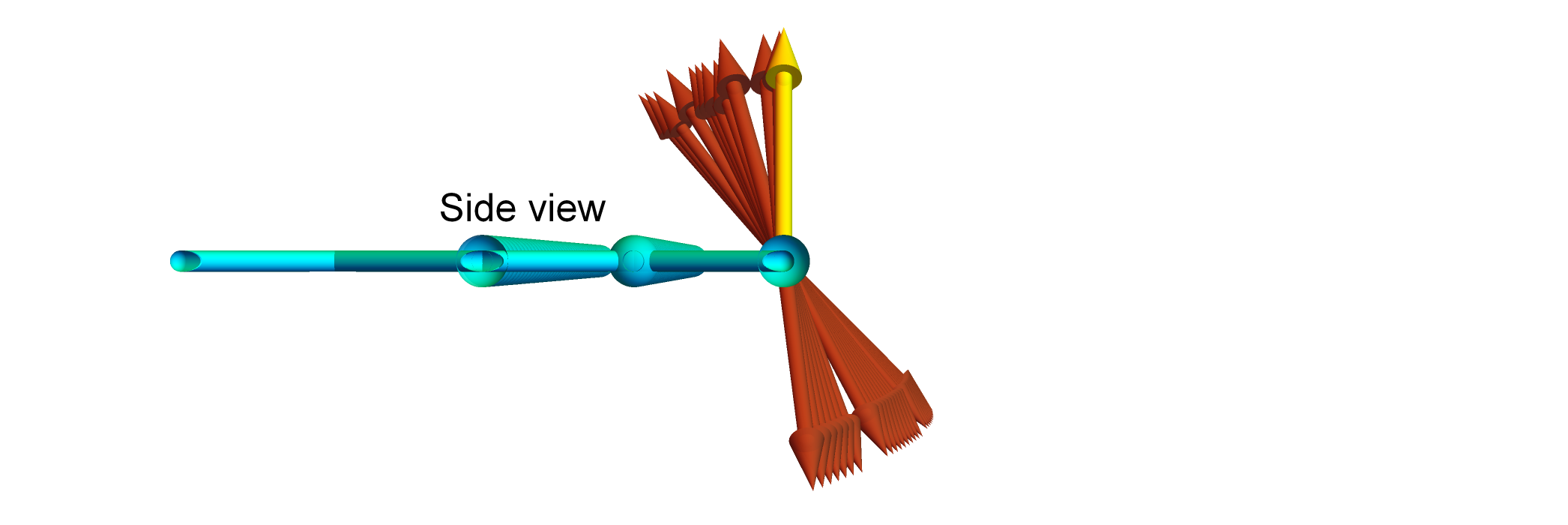}\includegraphics[width=0.5\textwidth]{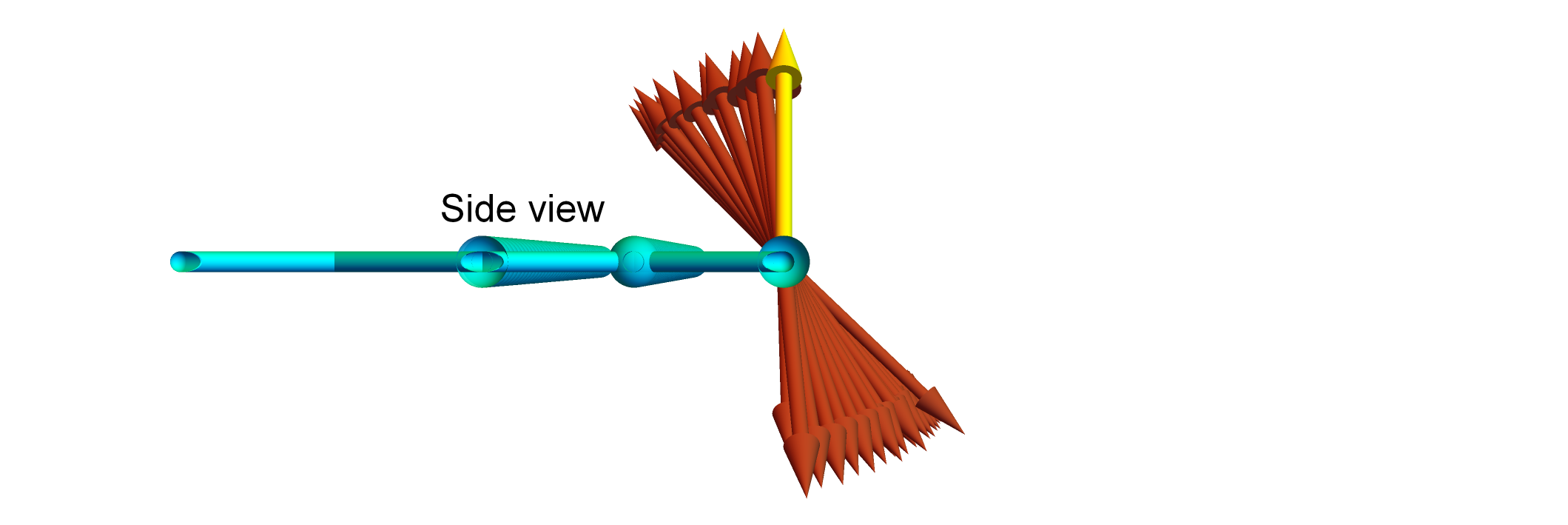}\\
	\caption{\label{fig3}RKKY interaction profiles shown as $J_{\beta \beta'}$ vs impurity separation $r/$a (a is the lattice constant), with $d_{z^2}$-orbital hybridization for both impurities, and for two different midgap energy levels (see figure\ \ref{fig2}): (a) 82 holes and (b) 22 holes. AFM (FM) alignment between both impurities is shown in light violet (light orange). The antinode structure of the wave function along the flake edge for states near the Fermi energy is seen to modulate the values of the different $J$'s. Insets show the log-log maxima of each antinode with separation, so that slopes give the decay exponent $d$ for the RKKY interaction, as shown by dotted lines. Lower panels represent different views of the classical ground state configuration for two impurities, considering the interactions shown in upper panels. The first impurity (yellow arrow) is fixed at $r/a=0$ with  arbitrary orientation, while the second impurity (brown arrow) is shown at every location $r/a\geq1$. The effect of the increasing DM interaction is clearly seen, especially in \emph{side} views, where the second impurity tends to twist towards the $xy$ plane as $r/a$ increases.}
\end{figure*}

In what follows, the positive (negative) values of these $J_{\beta\beta'}^{d_{z^2}d_{z^2}}$ correspond to AFM (FM) alignment between impurities (we will omit the superindex labels from now on).
Figure \ref{fig3} shows two RKKY interaction profiles, for Fermi levels fixed at two different \emph{p}-doped  levels in the optical gap, with 82 and 22 holes, respectively. These two Fermi levels are indicated in the inset of figure\ \ref{fig2}(b). The results show typical oscillating RKKY  behavior for all midgap energy levels.  The three components (Ising, XX, and DM) oscillate between AFM and FM behavior, and have different oscillation pattern for different doping levels, as one would expect (see below).
These oscillations are closely related to the $d_{z^2}$-orbital distribution along the flake edge of the wave function near the Fermi level, which shows similar spatial pattern of antinodes, as the characteristic Fermi wavelength dominates the resulting $J$. More importantly, the DM interaction is quite sizable, giving rise to a strong noncollinear interaction.

The Ising $J_{ZZ}$ and in-plane interactions $J_{XX}$ ($=J_{YY}$) both decay with impurity separation, with $J_{XX}$ decaying faster than $J_{ZZ}$.
The DM component $J_{DM}$ is seen to increase with distance, and reach larger amplitudes at the antinodes located after $r/a \simeq 10$.
The weaker $J_{DM}$ at short separation can be understood in terms of the finite size of the spin precession length induced by SOC, so that for close separations the host carrier scattering has not undergone significant spin rotation to generate a DM coupling.  Clearly, the onset for DM interaction
would depend on the strength of the SOC in the underlying band structure.

The strong DM interaction would be expected to yield a helical ground state for an assembly of multiple impurities.  This tendency is exemplified in the lower
panels of figure\ \ref{fig3}, where we schematically show the (classical) ground-state configuration of two impurities from different viewpoints (\emph{front, top} and \emph{side}).  The first impurity is shown as a yellow arrow, while the second is shown in brown.
The first impurity is fixed at an arbitrary orientation ($\theta_1=\pi/4$ and $\phi_1=\pi/2$), and the orientation of the second impurity that minimizes the energy at every $r/a$, is shown in the lower panels.  Taking as input the parameters of the RKKY interaction strength shown in the upper panels of figure\ \ref{fig3}, the lowest energy configuration for the second magnetic moment is given by $\tan (\phi_2 - \phi_1) = J_{DM}/J_{XX}$, and $\tan \theta_2 = J_{ZZ}^{-1} \sqrt {J_{DM}^2+ J_{XX}^2} \tan \theta_1$.  The orientation of the second impurity is seen to twist gradually as $r/a$ increases with respect to the first impurity, a clear manifestation of the effect of the increasing DM interaction. If DM were not present, either in an Ising+XX model or a Heisenberg model, the relative orientation of the impurities will remain in just one plane.

\begin{figure*}[ht]
	\centering \includegraphics[width=0.5\textwidth]{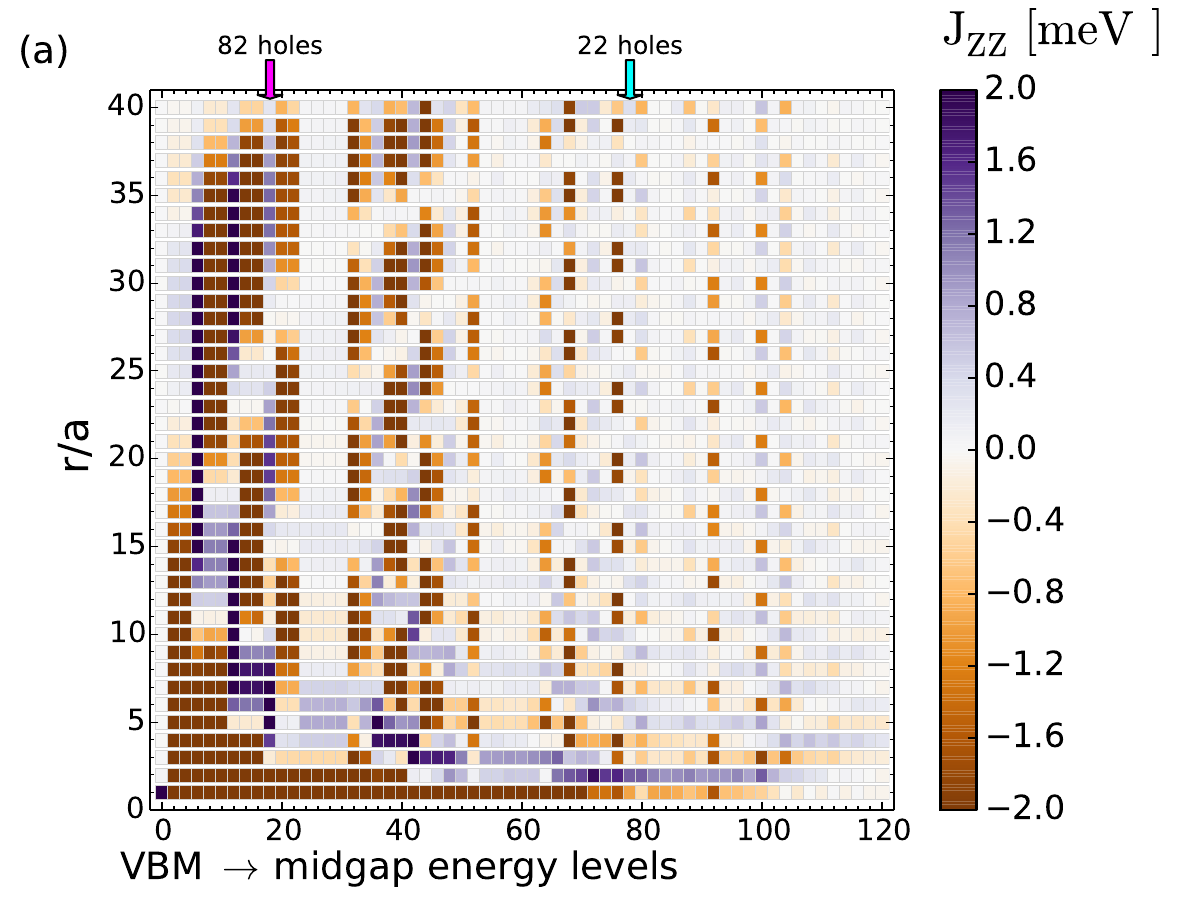}\includegraphics[width=0.5\textwidth]{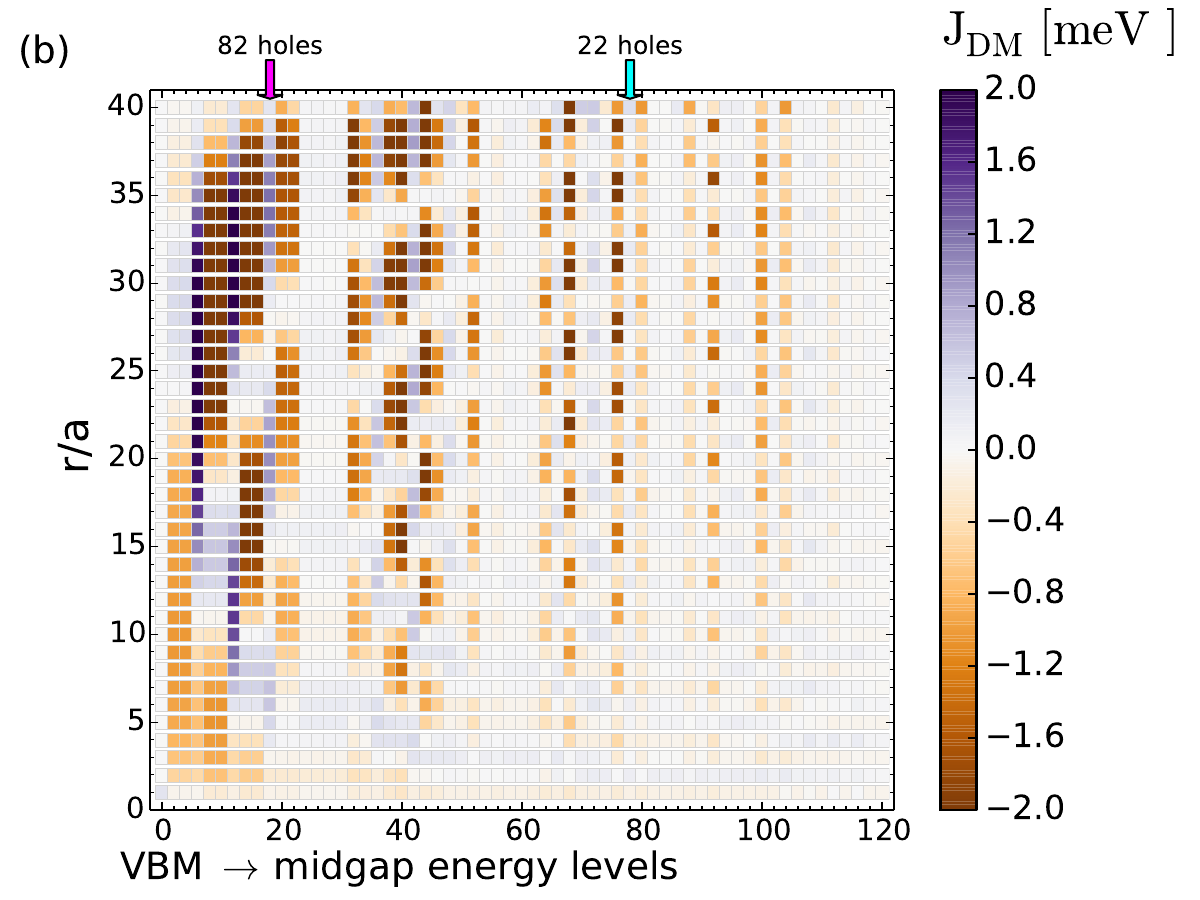}
	\caption{\label{fig4} (a) Ising and (b) Dzyaloshinskii-Moriya components of the RKKY interaction profiles, for various doping levels inside the optical gap, with orbital hybridization $d_{z^2}$ for both impurities. On the horizontal axis, zero corresponds to the VBM, and as the energy level index increases, one approaches the conduction band. Positive (violet) values correspond to AFM impurity alignment, while negative (orange) values correspond to FM impurity alignment. Doping levels in figure\ \ref{fig3} are indicated by magenta and cyan arrows at the top.}
\end{figure*}

We will now analyze the RKKY interaction for other midgap doping levels. Considering the same trajectory along the flake edge and $d_{z^2}$-orbital hybridization as before, figure\ \ref{fig4} shows two-dimensional maps of the Ising [panel (a)] and the DM [panel (b)] components of the RKKY indirect exchange, as a function of doping level, represented by the number of holes in the sample (horizontal axis), and as a function of the impurity separation $r/a$ (vertical axis). The two doping levels shown in figure\ \ref{fig3} (82 and 22 holes) are indicated at the top of the panels by magenta and cyan arrows, respectively.
We can observe that at all doping levels the interaction is oscillatory between FM (in orange) or AFM (in violet), changing sign along the trajectory, with  oscillation frequency that increases with higher Fermi energy (to the right on figure\ \ref{fig4}), associated with the shorter Fermi wavelength.
Notice also that as the Fermi level moves up towards the edge of the conduction band, the overall magnitude of the interaction decreases, although oscillations are still visible.  This drop in the effective exchange components is expected as the states higher in energy have a decreasing component of the $d_{z^2}$ orbital, have an overall lower density of states, and are increasingly more bulk-like--which moves the state into the bulk of the crystallite and away from the edges.
Regions of weak interaction, for doping such as between 24 and 30 (or 54 and 60), are the result of the first impurity being hybridized near a node of the $d_{z^2}$ orbital component at that Fermi level, suppressing the overall value of all $J$'s.
The in-plane $J_{XX}$ interaction terms (not shown) show the same qualitatively behavior as $J_{ZZ}$, but are seen to decay much more rapidly with separation, as illustrated by the cases shown in figure\ \ref{fig3}.  Indeed, for small separations ($r/a \lesssim 20$) the general trend is $J_{ZZ} \approx J_{XX} = J_{YY}$, and weak $J_{DM}$, signaling a Heisenberg-like interaction; but for larger separations ($r/a \gtrsim 20$), we see $|J_{XX}| \ll |J_{ZZ}|$, while $J_{DM}$ has become comparable to the Ising component. Hence, control of the doping levels throughout the optical gap provides a real tool for tuning the indirect exchange between impurities, being able to select non-collinear or isotropic, weak or strong, and AFM or FM behavior.

\begin{figure*}[ht]
	\centering
	\includegraphics[width=1\textwidth]{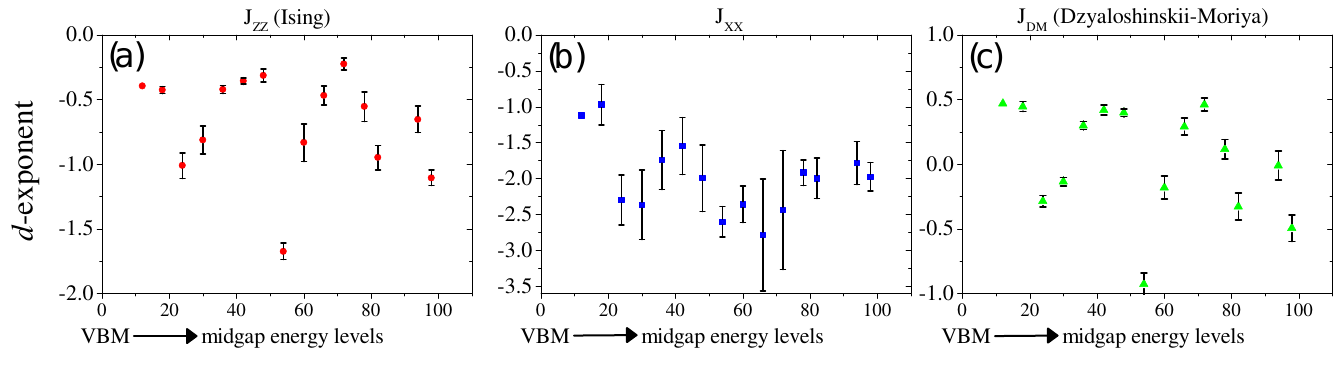}
	\caption{\label{fig5} RKKY decay exponent \emph{d} for midgap energy levels, $J \simeq r^d$. On the horizontal axis, zero corresponds to the VBM, and as the energy level index increases, one approaches the conduction band. Decay exponents for (a) Ising, (b) XX, and (c) Dzyaloshinskii-Moriya components. The vertical bars show the uncertainty when fitting the slopes.}
\end{figure*}

We now study the different spatial decay exponents for all the midgap doping levels.  In order to extract the decay exponent for each term of the effective interaction, $J_{\beta\beta'} \simeq r^d$, we plot the antinodes of each interaction curve vs impurity separation.  Examples of such plot, in a log-log scale are shown in the insets to each panel in figure\ \ref{fig3}, with a simple linear fit at large separations indicated by the dotted black lines.  Notice that both $J_{ZZ}$ and $J_{XX}$ have negative slopes in the insets, as expected, while $J_{DM}$ has a positive slope (reflecting the fact that this term increases initially with separation, as discussed above).
Figure\ \ref{fig5} shows the evolution of the envelope \emph{d}-exponent as a function of the midgap energy levels for each effective interaction term, $J_{\beta\beta'}\sim r^{d}$. For the Ising and XX interactions, $d$ has an ever more negative if slightly oscillating value for higher Fermi energy.  $J_{ZZ}$ has an exponent $d$ predominantly between $-1/2$ and $-1$, showing that a rather long-range Ising interaction is present for all midgap states.
In contrast, $J_{XX}$ is seen to quickly reach $d \simeq -2$, so that $J_{ZZ}$ clearly dominates at larger distances for all midgap dopings.
The DM $d$ value oscillates around zero, as that interaction appears to not have yet reached the fully asymptotic regime for these flake sizes.
It is also clear that the DM interaction is of the same general magnitude as the Ising component, giving rise to strong non-collinear relative alignment of impurities.

The unusual long range interaction between impurities reported here ($d \gtrsim -1$) can be seen to arise mainly from the  confinement of the edge/midgap states, which results in their effective 1D behavior, extended along the crystallite edge. This reduction in dimensionality is further aided by the finite size of the flake.  This is evident in the overall modulation of the states deeper in the gap, with an effective wavelength along the edge limited by the size of the flake.  It is remarkable that this modulation enhances the effective range of the indirect exchange interaction even for states closer to the conduction band (figure \ref{fig5}), although eventually $d \simeq -1$ for the Ising term. Notice that the envelope is further affected by the presence of antinodes in the edge states close to the Fermi level, as seen in figure\ \ref{fig4}.
Although we had seen evidence of unusually long-range interaction on edges of triangular MoS$_{2}$ flakes before  \cite{Avalos2016}, this was studied only for two midgap states, and we thought the effect fragile.  However, as we now see, the long range is rather robust to shifts in the Fermi level and for rather large flakes.  This should facilitate its experimental study without the need for fine tuning, and allow for simple gating STM studies, as reported recently \cite{Lu2014v2}.

We should also comment on the transferability of these results to other TMDs. We have  performed explicit calculations for WS$_{2}$, and found qualitatively similar results (not shown).  That is, we find slow decay of the envelope function--as well as oscillations--with the same general features as discussed for MoS$_{2}$. As the structure features of these two materials are rather similar, it is not surprising that the indirect exchange results agree even quantitatively for Ising and XX terms.  However, the stronger SOC coupling results in relatively larger DM terms. Other materials would exhibit the same qualitative behavior.

Finally, we note that the bulk/edge correspondence of the interaction has been previously reported in other 2D materials, such as graphene\cite{Black2010,Duffy2014} and silicine\cite{Zare2016}. For impurities on graphene nanoribbon edges, both zigzag\cite{Black2010} and armchair\cite{Duffy2014}, the decay has been found to be slower than $r^{-2}$ but only for small impurity separations, while for larger separations there is an exponential decay; for impurities interacting in the bulk of the nanoribbon the $r^{-2}$ decay is naturally recovered. Silicine shows a topological insulator phase, and when impurities sit on zigzag edges, the interaction decays as $r^{-1}$ and is much stronger than in the bulk \cite{Zare2016}. No unusually long range is reported for graphene or silicine, however. An interesting and different reduction of dimensionality in RKKY was also recently reported across graphene PN junctions \cite{Zhang2017VeselagoLens}.

\section{Conclusions}
We have studied the indirect interaction between magnetic impurities on the edges of a zigzag-terminated MoS$_{2}$ and WS$_{2}$ flakes. We focused on the role that edge states play on the interaction between impurities. We find that the 1D character of these states changes the dimensionality of the effective exchange interaction.
This gives rise, especially, to long ranged Ising interactions that could be tuned via the deposition of magnetic impurities on the flake edges, and the selection of midgap states through doping. All midgap states localized on the edges are shown to exhibit sub-2D decay behavior.  However, the in-plane component (XX) exhibits 2D behavior, especially at high doping.  The non-collinear Dyaloshinskii-Moriya interaction is seen to gradually rise with separation, reaching an amplitude comparable to the Ising component after a separation of nearly ten lattice constants.  All this behavior would suggest probing the tunable interaction between impurities near the edges of such TMD flakes, which would further result in stable  helical structures for many-impurity assemblies which could be examined by local probes, such as spin-polarized STM \cite{Zhou2010,Khajetoorians2016,Steinbrecher2017}. Given the electronic and structural similarity of several TMDs, stable and well defined phases in these systems may open the door to interesting spintronic applications in these versatile material systems. We hope
our predictions would motivate experiments with STM probes in the near future.

\ack We acknowledge support from NSF grant DMR 1508325. O. \'A.-O. acknowledges a research fellowship from the Condensed Matter and Surface Science program at Ohio University.

\section*{References}
\bibliographystyle{iopart-num}
\bibliography{AvalosOvandoEtAl}

\end{document}